\documentclass[12pt,a4]{article}
\usepackage{atbegshi}
\AtBeginDocument{\AtBeginShipoutNext{\AtBeginShipoutDiscard}}
\usepackage[affil-it]{authblk}
\usepackage{amsmath}%

\usepackage[titletoc,toc,title]{appendix}
\usepackage[ansinew]{inputenc}
\usepackage{graphicx}
\usepackage{amsmath}
\usepackage{amsthm}
\usepackage{bm}
\usepackage{layout}
\usepackage{float}
\usepackage{amsfonts}
\usepackage{amssymb}
\usepackage{txfonts}%
\usepackage{hyperref}
\newcommand{\unit}{1\!\!1}
\usepackage[font={small}]{caption}

\renewcommand{\thefootnote}{\fnsymbol{footnote}}

\newcommand\id{\leavevmode\hbox{\small1\kern-3.3pt\normalsize1}}

\begin{document}
\renewcommand{\thefootnote}{\arabic{footnote}}

\title{On the quantum measurement problem}


\author{{\v C}aslav Brukner$^{1,2}$}
 \thanks{\texttt{Submitted for the proceedings of the Conference Quantum UnSpeakables II: 50 Years of Bell's Theorem (Vienna, 19-22 June 2014)}}
\affil{$^1$Faculty of Physics, University of Vienna, Boltzmanngasse 5,
A-1090 Vienna, Austria.\\
$^2$Institute of Quantum Optics and Quantum Information, Austrian Academy of Sciences, Boltzmanngasse 3,
A-1090 Vienna, Austria.}

\date{\vspace{-5ex}}
  \maketitle
  \begin{abstract}

In this paper, I attempt a personal account of my understanding of the measurement problem in quantum mechanics, which has been largely in the tradition of the Copenhagen interpretation. I assume that (i) the quantum state is a representation of knowledge of a (real or hypothetical) observer relative to her experimental capabilities; (ii)  measurements have definite outcomes in the sense that only one outcome occurs; (iii) quantum theory is universal and the irreversibility of the measurement process is only ``for all practical purposes''. These assumptions are analyzed within quantum theory and their consistency is tested in Deutsch's version of the Wigner's friend gedanken experiment, where the friend reveals to Wigner whether she observes a definite outcome without revealing which outcome she observes. The view that holds the coexistence of the ``facts of the world'' common both for Wigner and his friend runs into the problem of the hidden variable program. The solution lies in understanding that ``facts'' can only exist relative to the observer. 

\end{abstract}

\setcounter{page}{1}
\subsection*{Two measurement problems}

There are at least two measurement problems in quantum mechanics\footnote{Two problems are assumed in Ref.~\cite{bub,pitowsky} and three problems are assumed in Ref.~\cite{maudlin}.}. The less prominent of the two (the ``small'' problem) is that of {\em explaining why a certain outcome -- as opposed to its alternatives -- occurs in a particular run of an experiment.} The bigger problem of the two (the ``big'' problem) is that of explaining the ways in which an experiment arrives at a particular outcome. It addresses the question of {\em what makes a measurement a measurement}\footnote{The proposed formulation of the two problems is inspired but not equivalent to the one of Ref.~\cite{bub,pitowsky} where the two categories of measurement problems were first introduced with the designations ``small'' and ``big''.}. 

In the following, I would like to present a personal account of my understanding of the measurement problems in quantum mechanics. My intention is not to argue that the approach I chose is the ``best'' way in any particular sense, but rather to demonstrate its logical consistency and to investigate what consequences the requirement for its consistency have for our understanding of physical reality. I will first present a probabilistic argument that explains why the measurement process is irreversible ``for all practical purposes''. Furthermore, by analyzing Deutsch's version of the Wigner's friend gedanken experiment, I will show that any attempt to assume that the measurement records (or ``facts'' or experiences) that coexist for both Wigner and his friend will run into the problems of the hidden variable program, for which I propose a Bell-type experiment. The conclusion is that these records can have meaning only relative to the observers; there are no ``facts of the world {\it per se}''.

Although I see my view of the quantum measurement problem broadly in the tradition of the Copenhagen interpretation, particularly within the information-theoretical approach to quantum mechanics~\cite{cbaz}, it contains elements from Qbism~\cite{qbism}, the relative interpretation of Rovelli~\cite{rovelli} and even the many-worlds interpretation. This indicates that the various interpretations might have much more in common than their supporters are willing to accept.

The solutions to the small measurement problem which have been offered to date basically present two underlying premisses. They either introduce ``hidden'' causes that determine which outcome will occur in a given experimental run (as in Bohm's hidden-variables theory), or they refute the basic notion of measurements resulting in definite outcomes (as in the Everett interpretation). None of that is really necessary. My position is that measurements have definite outcomes in the sense that only one outcome can be the result of a single experimental run. This is rather obvious. If it were otherwise, the notion of measurement would become ambiguous. If the outcome is not definitive, then no observation has occurred. This, however, does not exclude the possibility that the conditions that define a measurement are fulfilled for one observer but not for another. As far as discussions of the small measurement problem are concerned, I fail to see the reality of that problem. If one accepts the possibility of quantum probabilities being fundamentally irreducible, this problem vanishes. 

Let me explain that in more detail. Within quantum theory, a description of the quantum state of a system and of the measurement apparatus allows us to calculate the probability $p(a|x)$ to observe outcome $a$, for a measurement choice $x$. The probabilities are ``irreducible'' if there are no additional variables $\lambda$ in the theory, which potentially are yet to be discovered, such that when one conditions predictions on them, one has either $P(a|x,\lambda)=1$ or $0$, i.e. they allow the outcome to be predicted perfectly\footnote{The notion of irreducibility can be weakened to the requirement that the predictions conditioned on the variables are not more informative about the outcomes of future measurements than the predictions of quantum theory~\cite{colbeck}. Formally, for every measurement, the probability distribution conditioned on the variable cannot have lower (Shannon) entropy than the quantum probability distribution.}. Not only quantum mechanics, but every probabilistic theory in which probabilities are taken to be irreducible ``must have'' the small measurement problem. (The ``hypothetical collapse'' models~\cite{GRW,diosi,penrose} that predict the breakdown of quantum-mechanical laws on a macroscopic scale, if not supplemented by non-local hidden variables, also fall into this category.) The lack of the small measurement problem in the probabilistic theories would contradict the very idea of having irreducible probabilities~\footnote{The so-called ``non-local'' features of quantum theory are not a subject of the present article. I should, however, like to mention that once one accepts the notion that probabilities can be irreducible, there is no reason to restrict them to be {\it locally causal}~\cite{bell}, i.e. to be decomposable as: $p(a,b|x,y) = \int d\lambda \rho(\lambda) P(a|x,\lambda) P(b|y,\lambda)$, where $x$ and $y$ are choices of measurement settings in two separated laboratories, $a$ and $b$ are respective outcomes and $\rho(\lambda)$ is a probability distribution. It appears that the main misunderstanding associated with Bell's theorem stems from a failure to acknowledge the irreducibility of quantum probabilities {\em irrespectively} of the relative experimental space-time arrangements ~\cite{marek}. Bell's local causality accepts that probabilities for local outcomes can be irreducibly probabilistic, but requires those for correlations to be factorized into (a convex mixture of) probabilities for local outcomes. There is no need for imposing such a constraint on a probabilistic theory, where probabilities are considered to be fundamental. Rather, the notion of locality should be based on a operationally well-defined no-signaling condition, and it is this condition whose violation is at odds with special relativity.}.

The big measurement problem is more subtle. It can be illustrated by the following situation. As students, we are taught that there are two processes a quantum state can undergo. First, the deterministic, unitary, and continuous time evolution of the state (of a system, possibly together with its environment) that obeys the Schroedinger equation or its relativistic counterpart. Second, the probabilistic, non-unitary, and discontinuous change of the state, called ``projection'' or ``collapse'', brought about by measurement. Equipped with this knowledge, we attend a practical training in a quantum optics laboratory, where we see various pieces of equipment, such as photonic sources, beam-splitters, optical fibers, mirrors, photodiodes, phase shifters etc., for the first time. The instructor sets us the task of computing the evolution of the photonic quantum state in the set-up prepared on the optical table. We soon realize that we are in trouble. There is nothing in the theory to tell us which device in the laboratory corresponds to a unitary transformation and which to a projection! We start to ask questions. What makes a photodiode a good detector for photons? And why is a beam splitter a bad detector? At least manufacturers of photon detectors should know the answers to these questions, shouldn't they? Or perhaps the measurement is not completed in the detector, but only when the result is finally recorded in a computer, or even in the observer's mind? Bell sardonically commented~\cite{bell}: ``What exactly qualifies some physical systems to play the role of 'measurer'? Was the wave function of the world waiting to jump for thousands of millions of years until a single-celled living creature appeared? Or did it have to wait a little longer, for some better qualified system ... with a Ph.D.?'' 

One possibility to address these questions would be to dismiss the big measurement problem as a pseudo-issue, just as we did for the small one. If quantum theory is understood as a fundamental theory of observations and observers' actions upon these observations, then measurement can be introduced as a primitive notion, which cannot be subject to a complete  analysis, not even in principle. At most, one could motivate it informally, through an appeal to intuition and everyday experience. It seems to me that this path is taken by some proponents of the Copenhagen interpretation and Quantum Bayesianism (QBists), for example when Fuchs and Scheck write~\cite{fuchs} ``a measurement is an action an agent takes to elicit an experience. The measurement outcome is the experience so elicited''. Such a view is consistent and self-contained, but in my opinion, it is {\it not} the whole story. It is silent about the question: what makes a photon counter a better device for detecting photons than a beam splitter? Yet the question is scientifically well posed and has an unambiguous answer (which manufacturers of photodetectors do know!).  

I would like to express clearly that I do agree with the Qbists and the Copenhagenists on the necessity of a {\it functional} distinction between the object and the subject of observation. This distinction is at the heart of Bohr's epistemological argument that measurement instruments lie outside the domain of the theory, insofar as they serve their purpose of acquiring empirical knowledge. Regretfully, this argument has repeatedly been misinterpreted in textbooks and articles and ``replaced by the crude physical assumption that macroscopic systems behave classically, which would introduce an artificial split of the physical world into a quantum microcosms and a classical macrocosms.''~\cite{osnaghi}. The ``cut'' is not between the macro and micro worlds but between the measuring apparatus and the observed quantum system. It is of epistemic, not of ontic origin. 

Bohr and Heisenberg seem to have disagreed about the movability of the cut~\cite{max}. As Heisenberg recalls in his letter to Heelan~\cite{heelan} (quoted in Ref.~\cite{max}): ``I argued that a cut could be moved around to some extent while Bohr preferred to think that the position is uniquely defined in every experiment''. In my understanding, the two views are not conflicting and can be brought into accordance. Heisenberg acknowledges the universality of the laws of quantum mechanics in the sense that every system, including the measuring instrument, is {\it in principle} subject to these laws. Of course, in moving the cut, the measurement instrument loses its function as a means for acquiring knowledge about a quantum system but becomes itself a quantum system -- an object that can be observed by a further set of measurement instruments. Bohr, however, believes ``that for a {\it given} (my italics) experimental setting the cut is determined by the nature of the problem ...'', as he writes in a 1935 letter to Heisenberg~\cite{AHQP} (quoted in Ref.~\cite{max}). The cut is hence movable in principle, but is fixed in any concrete experimental set-up. Still, we might wonder what fixes the position of the object--instrument cut in a concrete experimental set-up? Here, Bohr encounters the big measurement problem. 

The question of the meaning of the quantum state is closely related to the measurement problem(s). Which approach one takes in addressing the later depends on the specific view one has on the former. The next section is devoted to this question. 

\subsection*{What is the quantum state?}

The discussion over the meaning of quantum states is often presented as a conflict between two fundamentally opposed approaches. The first approach speaks of ``states of reality'' that are independent of any empirical access, and implicitly assumes the existence of such states. The second approach refers to observations, and what we can know about them and deduce from them. This approach requires differentiation with regard to the question ``Knowledge about what?''. Insofar as the quantum state is seen as representing the observer's incomplete knowledge about an assumed ``state of reality'', it is not fundamentally dissimilar to the first approach. This is why, to use modern terminology, the distinction between a realist interpretation of a quantum state that is ``psi-ontic'' and one that is ``psi-epistemic''~\cite{spekkens} -- which actually is a distinction between two kinds of hidden variable theory -- is only relevant to supporters of the first approach. 

An alternative exists. The quantum state can be seen as a mathematical representation of what the observer has to know in order to calculate probabilities for outcomes of measurements following a specific preparation. However, one could also object to this ``operationalist's view''. Malin phrased it nicely~\cite{malin}: ``What if the knower is a physicist who had a martini before trying to 'know'? What if a person who knows just a little physics learns of the result? What if he had a martini? Somehow we feel that such questions are irrelevant.'' He continues: ``To avoid difficulties of this kind regarding the epistemic interpretation, we can consider a quantum state as representing not actual knowledge (which requires a knower), but the available or potential knowledge about a system.''  

Of course, the argument that quantum theory does not apply in the absence of observers has not been made. Yet when calculating a quantum state, it might help to think of a {\it hypothetical} observer for whom the quantum state stands for her knowledge~\footnote{Peres correctly notes that considering hypothetical observers is not a prerogative of quantum theory~\cite{peres}. They are also used in thermodynamics, when we say that a perpetual-motion machine of the second kind cannot be built, or in the theory of special relativity, when we say that no signal can be transferred faster than the speed of light.}. For example, when quantum cosmologists talk about the pressure of a primordial state of the universe, we can make sense of it if we imagine a well-defined experimental procedure a  hypothetical observer could apply on the state to provide an operational meaning to the term ``pressure''. The ultimate meaning is presented by current cosmological observations, based on which we reconstruct the idea of the early universe's pressure. (The observer here is always considered to be external to the universe. The ``wave function of the universe'' that would include the observer is a problematic concept, as it negates the necessity of the object--subject cut.). This is compatible with Malin's view~\cite{malin} that ``quantum states represent the available knowledge about the potentialities of a quantum system, knowledge from the perspective of a particular location in space'', not of any actual observer. 

I share Malin's view on the meaning of the quantum state, which is essentially the one supported by Copenhagenists and Qbists. I would like to add just one, but an important, aspect to this view: 
{\em The quantum state is a representation of knowledge necessary for a hypothetical  observer -- respecting her experimental capabilities -- to compute probabilities of outcomes of all possible future experiments}. An explicit reference to the observer's experimental capabilities is crucial to address the big measurement problem. The ``knowledge'' here refers to Wigner's definition of the quantum state~\cite{wigner}: ``... the state vector is only a shorthand expression of that part of our information concerning the past of the system which is relevant for predicting (as far as possible) the future behaviour thereof.''

The available experimental precision will in every particular arrangement determine to which objects the observer can meaningfully assign quantum states. This agrees with Bohr's view ``that for a given experimental setting the cut is determined by the nature of the problem ...'' That there is nothing in the theory that would prohibit reaching the necessary experimental precision to allow a meaningful state assignment to objects of increasingly large sizes -- eventually as large as our measurement devices -- reflects Heisenberg' view that the ``cut can be shifted arbitrarily far in the direction of the observer'', as he wrote in an unpublished paper~\cite{heisenberg} from 1935, in which he outlined his response to the Einstein, Podolsky, and Rosen paper from the same year\footnote{In the same paper, Heisenberg concludes: ``... the quantum mechanical predictions about the outcome of an arbitrary experiment are independent of the location of the cut ...'' This can be seen as a consequence of ``purification'' in quantum theory, which states that every mixed state of system A can always be seen as a state belonging to a part of a composite system AB that itself is in a pure state. This state is unique up to a reversible transformation on B. The assumption of purification is one of the central features of quantum theory, which, taken as an axiom together with a few other axioms, makes it possible to explain why the theory has the very mathematical structure it does~\cite{chiribella}.}. The measurement instrument and the observer can be included in the quantum mechanical description, and then observed by someone else, a ``superobserver'', for whom the original measurement instrument loses its previous status as a means for acquiring knowledge. For this purpose, she needs another set of ``superinstruments'' that are superior to the original instruments regarding their precision. 

\subsection*{FAPP  irreversibility}
 
The distinct outcomes of a measurement apparatus are associated with ``macroscopically distinct states''. (Only in that aspect does ``macroscopicity'' play a role in the measurement process.) These are defined as states that can still be differentiated even in those cases where the measurements are imprecise and coarse-grained. If, for example, just a few spins of a large magnet are flipped, then the entire quantum state of the magnet will change into an orthogonal one. Yet, at our macroscopic level, we will still perceive it as the very same magnet. In order for the change to become noticeable even in a coarse-grained measurement, the quantum states of a sufficiently large number of spins need to be changed.   

In order to quantify the distinguishability of such macroscopic states, consider a spin-$j$ system, with $ j \gg 1$, and the coarse-grained measurements with the POVM elements
\begin{equation}
P_{\Omega_0}=\frac{2j+1}{4\pi} \iint_{\Omega_0} d\Omega |\Omega\rangle \langle \Omega|
\end{equation}
as a model example, where $|\Omega\rangle= \sum_{m=-j}^{+j} {2j \choose j+m}^{1/2} \cos^{j+m}\frac{\theta}{2} \sin^{j-m}\frac{\theta}{2} e^{-im\phi_0} |m\rangle $ is the spin coherent state and $\theta$ and $\phi$ are the polar and azimuthal angles respectively, corresponding to the solid angle $\Omega$. The size of the integration region around the solid angle $\Omega_0=(\theta_0, \phi_0)$ is taken to be such that its projection $\Delta m$ along the $z$ axis is much larger than the intrinsic uncertainty of the coherent states\footnote{When we introduce coarse-grained observables, we need to define the states that are ``close'' to each other to conflate them into coarse-grained outcomes. However, the terms ``close'' or ``distant'' make sense in a classical context only. There, ``close'' states correspond to neighboring outcomes in the real configuration space. For example, the coherent states conflated in the single outcome $\Omega_0$ of the POVM all correspond to approximately the same direction $\Omega_0$ in real space. Therefore, certain features of classicality need to be presumed before macroscopic states can be defined. An alternative would be the attempt to reconstruct the notions of closeness, distance, and space -- and consequently, also the theories referring to these notions, such as quantum field theory -- from within the formalism of the Hilbert space only. Useful tools for this attempt might be preferred tensor factorizations, coarse-grained observables, and symmetries. The results of Refs.~\cite{mueller,dakic} present first progress towards this goal. The most elementary quantum system, the qubit, resides in an abstract state space with SU(2) symmetry. This is locally isomorphic to the group SO(3) of rotations in three-dimensional space. Considering  directional degrees of freedom (spin), this symmetry is found to be operationally justified in the symmetry of the configuration of macroscopic instruments used for transforming the spin state. Hereby one assumes that quantum theory is ``closed'': the macroscopic instruments do not lie outside of the theory, but are described from within it in the limit of a large number of its constituents (as coherent states or ``classical fields'')~\cite{dakic}.}, $\Delta m \gg \sqrt{j}$. Under the coarse grained measurement, any state $\hat{\rho}$ can effectively be described in terms of a positive probability distribution (the well-known Q-function)~\cite{kofler}.
\begin{equation}
Q(\Omega) = \frac{2j+1}{4\pi} \langle\Omega|\hat{\rho}|\Omega\rangle.
\end{equation}
Specifically, the probabilities for the POVM outcomes can be obtained by averaging the Q-function: $P_{\Omega_0} = \iint_{\Omega_0} Q(\Omega)$. Hence, the description in terms of $Q(\Omega)$ is effectively classical and it leads to the classical limit of quantum mechanics\footnote{The classical world arises from within quantum theory when neighboring outcomes are not distinguished but bunched together into slots in the measurements of limited precision. What would the classical world look like if {\it non-neighboring} outcomes were conflated to slots? To address this question, one could imagine an experiment on a person whose nerve fibers behind the retina are disconnected and again reconnected at different, {\it randomly} chosen, nerve extensions connecting to the brain. It seems reasonable to assume that the neighboring points of the object that is illuminated with light and observed by the person's eye will no more be perceived by the person as neighboring points. One may wonder if, in the course of further interaction with the environment, the person's brain will start to make sense out of the seen ``disordered classical world'', or if it will post-process the signals to search for more ``ordered'' structures as a prerequisite for making sense out of them. The latter may eventually nullify the effect of the random reconnection of nerves, and the person will again perceive the ordinary classical world.}~\cite{kofler}. Since $Q(\Omega)$ represents a complete description of the system under coarse-grained measurements, I will call it the ``macroscopic state''. This approach to classicality differs conceptually from and is complementary to the decoherence program that is dynamical and describes correlations of the system with other degrees of freedom which are integrated out~\cite{zurek}.

A measure of the distinguishability between two probability distributions $P(\Omega)$ and $Q(\Omega)$ is the Euclidean scalar product $({\mathbf P},{\mathbf Q}) := \iint d\Omega \sqrt{ P(\Omega) Q(\Omega)}$. If two probability distributions are perfectly distinguishable $({\mathbf P},{\mathbf Q})= 0$, while if they are identical $({\mathbf P},{\mathbf Q})=1$. Consider two pure quantum states $|\psi_1\rangle$ and $|\psi_2\rangle$ with the Q-functions $|\langle \psi_1 |\Omega\rangle |^2$ and $|\langle \psi_2 |\Omega\rangle |^2$, respectively. Then $ \frac{2j+1}{4\pi} \iint d\Omega \left| \langle \psi_1|\Omega\rangle| |\langle \Omega \right|\psi_2| \geq \frac{2j+1}{4\pi} |\iint d\Omega \langle \psi_1 |\Omega\rangle \langle \Omega|\psi_2 \rangle| =|\langle \psi_1|\psi_2 \rangle|$, where $\frac{2j+1}{4\pi}\iint d\Omega |\Omega\rangle \langle\Omega \rangle =\unit$. This shows that distinguishability between the quantum states and distinguishability between macroscopic states are two different notions. The latter implies the former, but the opposite is not true. Say, spin $j$ is composed of $N$ spins-1/2. In this case, a number of spins of the order of $\sqrt{N}$ need to be flipped in order to arrive at a macroscopically distinct state. Only then do we perceive it as a new state of magnetization. It is now a new ``fact''.

The macroscopic states are robust. This means that they are stable against perturbations, which may for example be caused by repeated coarse-grained observations. In other words, the Q-function before and after a coarse-grained measurement is approximately the same~\cite{kofler2}. It therefore becomes possible for different observers to repeatedly observe the same macroscopic state. The result is a certain level of intersubjectivity among them. If we assume, however, that quantum mechanics is universally valid, then it is in principle possible to undo the entire measurement process. Imagine a superobserver who has full control over the degrees of freedom of the measuring apparatus. Such a superobserver would be able to decorrelate the apparatus from the measured system. In this process, the information about the measurement result would be erased. Seen from this perspective, ``irreversibility'' in the quantum measurement process merely stands for the fact that it is extremely difficult -- but not impossible! -- to reverse the process. It is irreversible ``for all practical purposes'' (or ``FAPP,'' to use Bell's acronym). 

I have often heard the following objection to FAPP: No matter how low the probability is to reverse the evolution in the measurement process, it is still there. How is it possible to settle the question of what actually exists by an approximation? In my eyes, such questions do not take into consideration the simple fact that quantum theory cannot be both, universal and not irreversible merely FAPP. While on the one hand, measurements have to result in irreversible facts (otherwise, the notion of measurement itself would become meaningless, as no measurement would ever be conclusive), this irreversibility on the other hand must be merely FAPP if quantum theory is in principle applicable to any system. Any system means that the measuring apparatus itself can also be subject to the laws of quantum theory. My main point is the following. While it is obviously possible to describe the subject as an object, it then has to be the object for another subject. In my eyes, not enough thought has gone into the fundamental nature of FAPP. More research on the philosophy of FAPP, if you like, should be done by philosophers of physics. This, in my eyes, would contribute to the resolution of the problem in a much deeper way than the perpetual attempts to expel this term from the foundations of physics based on presupposed philosophical doctrine. 

Detection devices, such as photographic plates or photo-diodes, consist of a large number of constituents in a certain ``metastable state''. Their interaction with the observed quantum systems brings them into a ``stable state'' that can be distinguished from the initial one even under coarse-grained observations. This transition is signified by the ``click'' in the detector or a new position of the pointer label. In both the metastable and the stable state, the constituents of the instrument can be in any of a large number of quantum states that correspond to the respective macroscopic states. In order to understand how irreversibility FAPP is possible, it is crucial to realize that not only the initial and final quantum states of the instrument are imprecisely known, but also the full details of the interactions (i.e. Hamiltonian) among its constituents and with the environment. Even if it were possible to know the initial and final states precisely, the lack of precise knowledge of these interactions prevents us from reversing the measurement process. Moreover, a photodetector does not spontaneously ``de-click''. It does not turn itself back into the initial metastable state and and it does not emit the photon into its initial state. 

The irreversibility of the measurement process might be explained in quantum mechanical terms, but as metastable and stable states of the detector are macroscopic states, a classical explanation of irreversibility is sufficient. In fact, nothing ``quantum'' is indispensable for ``solving'' the big measurement problem. The ``problem'' is essentially present in classical measurements as well.

For classical (chaotic) systems, the physical state after a certain time can become unpredictable if the solutions of dynamical equations are highly sensitive, either to initial states or to uncontrollable external perturbations of the Hamiltonian. The situation is different in quantum mechanics because of the unitarity of the dynamical evolution: the scalar product between the unperturbed and the perturbed state is constant such that an uncertainty in initial states will not grow in time. However, an uncontrollable external perturbation to the Hamiltonian can explain FAPP irreversibility for both the classical and quantum case. Below I consider one such model. 

Consider the detection device to be a classical dynamical system for which the state $\mathbf{x}_t$ at an arbitrary time $t$ is given in terms of a flow $\mathbf{x}_t = f_t(\mathbf{x}_0)$ on the initial state $\mathbf{x}_0 = (\mathbf{q},\mathbf{p})$, where $\mathbf{q}$ are the positions and $\mathbf{p}$ the momenta of all the system's constituents. The flow is assumed to be reversible, i.e. there exists the involution $\pi(\mathbf{q},\mathbf{p})=(\mathbf{q},-\mathbf{p})$ with $\pi^2=\unit$ for which for all times $\pi f_t \pi =f^{-1}_t$. We now choose two regions $A$ and $B$ of the phase space. We assume a uniform probability distribution of the state over $A$ to exist at the outset. The probability of finding the state in the set $B$ at the time $t$ when, at the time $t = 0$, the state was in the set $A$ is given by the volume fraction of the states from $A$ that evolves in $B$.
\begin{equation}
\mbox{Prob}[\mathbf{x}_t \in B| \mathbf{x}_0 \in A] = \frac{|A \cap f^{-1}_t B|}{|A|}. \label{ris}
\end{equation}
Here, $|X|$ is the Lebesgue measure of set $X$. In the remaining argument I will assume that $B = f_t A$, for which the probability~(\ref{ris}) is 1. 

Suppose now that we want to reverse the evolution, but do not have precise control of the flow $f_t$, for example due to uncontrollable influences from the environment. Hence, the inverse flow $f'_t \neq f_t$ is perturbed, where we assume $\pi f'_t \pi =f'^{-1}_t$.
At the time $t$, we inverse all the momenta and set the time again to 0 for simplicity. (Note that inverting momenta does not require measuring them, which would be impossible due to the limited precision of the instruments, nor does it require to know them precisely. An arrangement with elastic bounce of the molecules would be sufficient~\cite{peres2}. If $f'_t=f_t$, it would be sufficient to inverse the momenta to perfectly reverse the evolution.) Then, the probability to find the state in the set $\pi A$ at the time $t$ when, at time $t=0$, the state was in the set $\pi B= \pi f_tA$ is given by
\begin{equation}
\mbox{Prob}[\mathbf{x}_t \in \pi A| \mathbf{x}_0 \in \pi f_t A] = \frac{|\pi f_t A \cap f'^{-1}_t \pi A|}{|\pi f_t A|} = \frac{|f_t A \cap f'_t A|}{|A|},\label{prob}
\end{equation}
where we used $|\pi X|=|X|$ and the Liouville theorem $|f_t X|=|X|$. (Once one arrives at the states from the set $\pi A$, one can obtain those from the initial set $A$ by simply inverting the momenta. This might induce additional imprecisions, neglected here for simplicity.) The expression~(\ref{prob}) has an operational meaning, namely that of the ``probability for reversing the evolution''. It is the probability to find the system in the initial state under first the forward, and then the reverse, perturbed, flow.

\begin{figure}
\begin{center}
\includegraphics[width=9.8cm]{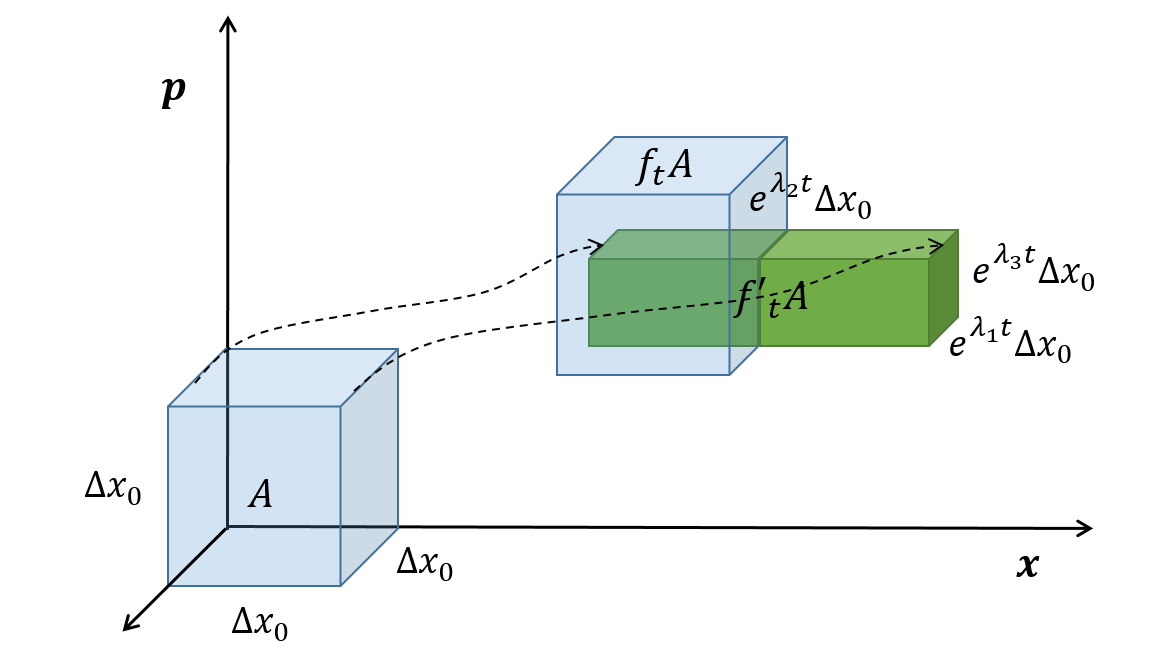}
\end{center}
\caption{Schematic illustration of the phase space evolution in both the regular and the chaotic case. An infinitesimal element $A$ of volume $(\Delta x_0)^n$, where $n$ is the dimension of the phase space, evolves in $f_tA$ under a regular flow or in $f'_tA$ under a chaotic flow. The volume of the overlap between the two evolved elements has the linear size of $\Delta x_0$ along all directions of divergence and $e^{\lambda_i t} \Delta x_0$ $(\lambda_i <0)$ for every direction $i$ of contraction.} \label{figure1}
\end{figure}

The classical explanation of irreversibility is based on the notions of mixing and coarse graining. Mixing is a property of chaotic systems for which at least one of the Lyapunov exponents is positive. (If the system is a Hamiltonian system, the sum of all Lyapunov exponents is zero. If the system is dissipative, the sum is negative.) Two trajectories in phase space with an initial separation $\Delta x_0$ along dimension $i$ diverge at a rate given by $ \Delta x_t  \approx e^{\lambda_i t}  \Delta x_0 $, where $\lambda_i>0$ is a Lyapunov exponent. Suppose that $|A|=(\Delta x_0)^n$ corresponds to the small volume that can still be distinguished from other such volumes in a coarse-grained observation, and $n$ is the dimension of the phase space. Furthermore, suppose that $f_t$ is regular and does not significantly change the form of $A$, while $f'_t$ is chaotic. One has for the phase space volume $|f'_t A| \approx e^{\sum_i^n \lambda_i} (\Delta x_0)^n$. Then the probability~(\ref{prob}) is specified by the volume $|f_t A \cap f'_t A|$ of the largest overlap between the volume elements $f_t A$ and $f'_t A$. This volume has the linear size $\Delta x_0$ along all directions of stretching  and $ e^{\lambda_i t} \Delta x_0$ along every direction $i$ of contraction. Hence, the probability is bounded as
\begin{equation}
\mbox{Prob}[\mathbf{x}_t \in \pi A| \mathbf{x}_0 \in \pi f_t A] \leq e^{-\sum'_{i} \lambda_it}, 
\end{equation}
where the sum $\sum'_i$ is over positive Lyapunov exponents. The probability of reversing the evolution and arriving at the initial state is negligibly small after several multiples of the characteristic time $t \sim 1/\sum'_i \lambda_i$. The above argument can explain the classical irreversibility of macroscopic states in detection instruments, but for completeness, I will below present a quantum version for it.

One can define a quantum mechanical measure of the state revival when an imperfect time-reversal evolution is applied to a quantum system. We will illustrate this with an example of a spin system. Suppose that an initial quantum state $|\psi\rangle$ evolves during a time $t$ under a Hamiltonian $H_0$ into the final state $|\psi(t)\rangle$. The two states define macroscopic states (Q-functions) $P(\Omega,0)=|\langle \Omega | \psi \rangle|^2$ and $P(\Omega,t)=|\langle \Omega | \psi(t) \rangle|^2$, respectively. Any attempt to reverse the evolution and arrive back at the initial macroscopic state will result in an application of a perturbed, slightly different Hamiltonian $\hat{H}=\hat{H}_0+\hat{V}$ with perturbation $\hat{V}$. Perfect recovery of macroscopic state could be achieved only if one could have a sufficient control over $\hat{H}$. In a realistic situation, however, such a control is FAPP impossible. 

As a measure of reversibility we use the scalar product
\begin{equation}
({\mathbf P},{\mathbf Q}) =\iint d\Omega \sqrt{P(\Omega,0)Q(\Omega,t)}
\end{equation}
between the probability distribution $P(\Omega,0)= |\langle \Omega|\psi \rangle|^2$ of finding initially the system in a macroscopic ``phase point'' $\Omega$ and the probability distribution
\begin{equation}
Q(\Omega,t)= |\langle \Omega |e^{\frac{i\hat{H}t}{\hbar}}e^{\frac{-i\hat{H}_0t}{\hbar}}|\psi \rangle|^2
\end{equation}
of finding it there after a combined evolution: forward evolution in duration of $t$  under the Hamiltonian $\hat{H}_0$ and then backward evolution in duration of $t$ under $-\hat{H}$.
The combined evolution embodies the notion of time-reversal. If for some $t > 0$, one has $({\mathbf P},{\mathbf Q}) \approx 1$, the evolution is reversed at the macroscopic level. (Note that the reversed quantum state $e^{\frac{i\hat{H}t}{\hbar}}e^{\frac{-i\hat{H}_0t}{\hbar}}|\psi \rangle$ does not need to be identical to the initial one $|\psi\rangle$ to have reversibility at the macroscopic level. It is only important that they approximately correspond to the same macroscopic state.) 

Consider for simplicity a non-degenerative Hamiltonian $\hat{H}_0$ with eigenstates $|\alpha_0\rangle$ for eigenvalues $E^0_\alpha$, and the perturbed Hamiltonian $\hat{H}$ with eigenstates $|\alpha \rangle$ for eigenvalues $E_\alpha $. For simplicity, I assume an extremely weak perturbation $\hat{V}$ for which $E_\alpha=E^0_\alpha + \langle \alpha_0 |\hat{V} |\alpha_0 \rangle$ 
and $\langle \alpha | \beta_0\rangle=\delta_{\alpha,\beta_0}$. 
Expanding $|\psi\rangle=\sum_\alpha \psi_\alpha |\alpha\rangle$, one has
\begin{equation}
Q(\Omega, t)= \sum_{\alpha,\beta} \psi_\alpha \psi^{*}_\beta \langle \Omega |\alpha  \rangle \langle \beta|\Omega \rangle e^{\frac{i (V_\alpha - V_\beta) t}{\hbar}}, \label{rmt}
\end{equation}
where $V_\alpha:=\langle \alpha |\hat{V}|\alpha \rangle$. 

The value of~(\ref{rmt}) depends on the statistical distribution of $V_\alpha-V_\beta$ over different perturbations~\cite{peres3}. This means that in every repetition of our procedure the system will be differently perturbed during its backwards evolution. For chaotic systems, one assumes that this distribution is governed by a random matrix theory~\cite{chaos}. According to this theory,  $V_\alpha$ are independent random numbers, and for a large number of eigenstates, the distribution can be approximated by a Gaussian one $g(V_\alpha)=\frac{1}{\sqrt{\pi}\sigma} e^{-\frac{(V_\alpha-W_\alpha)^2}{\sigma^2}}$ around the mean value $W_\alpha$. Taking an ensemble average over different perturbations, one obtains $\langle e^{\frac{iV_\alpha t}{\hbar}} \rangle_{pert} = \int^\infty_{-\infty} dx g(x) e^{\frac{ixt}{\hbar}} e^{\frac{iW_\alpha t}{\hbar}} = e^{\frac{iW_\alpha t}{\hbar}} e^{-\frac{\sigma^2 t^2}{4\hbar^2}}$. In the model the distribution spread $\sigma$ is taken to be much smaller than the level spacing to ensure no correlations in the distribution, $\langle e^{\frac{i(V_\alpha-V_\beta) t}{\hbar}} \rangle_{pert} = \langle e^{\frac{iV_\alpha t}{\hbar}} \rangle_{pert}\langle e^{\frac{-iV_\beta t}{\hbar}} \rangle_{pert}$. Finally, we obtain
\begin{equation}
\langle Q(\Omega,t) \rangle_{pert}= \sum_{\alpha,\beta} \psi_\alpha \psi^{*}_\beta \langle \Omega |\alpha  \rangle \langle \beta|\Omega \rangle e^{\frac{i (W_\alpha - W_\beta) t}{\hbar}}e^{-\frac{\sigma^2 t^2}{2\hbar^2}}=|\langle\Omega|\phi(t)|^2 e^{-\frac{\sigma^2 t^2}{2\hbar^2}}, 
\end{equation}
where in the final step we introduce $\phi_\alpha (t):= \psi_\alpha e^{\frac{iW_\alpha t}{\hbar}}$ and $ |\phi(t)\rangle:=\sum_\alpha \phi_\alpha(t) |\alpha\rangle$.

Using $\langle \sqrt{Q(\Omega,t)} \rangle_{pert} \leq \sqrt{\langle Q(\Omega,t) \rangle_{pert}}$ one obtains for the measure of reversibility
\begin{equation}
\langle({\mathbf P},{\mathbf Q})\rangle_{pert} \leq \iint d\Omega 
|\langle \Omega |\psi\rangle| |\langle \Omega| \phi (t) \rangle| e^{-\frac{\sigma^2 t^2}{4\hbar^2}} \leq e^{-\frac{\sigma^2 t^2}{4\hbar^2}}.
\end{equation}
We see that for random perturbations on average the macroscopic state will significantly change after first forward evolution in duration of time $t$, and then reverse evolution in duration of time $t$, if $t > \hbar/\sigma$, indicating FAPP irreversibility. Specifically, for $t\rightarrow \infty$, one has $\langle({\mathbf P},{\mathbf Q})\rangle_{pert} \rightarrow 0$. The regime beyond the validity of weak perturbation can be treated using the results from the field of quantum chaos~\cite{jacquod}.

For the present distribution over perturbations both the evolution of the macroscopic state and of the quantum state is FAPP irreversible. One can find, however, such distributions for which the evolution of the macroscopic state is reversible and the evolution of the quantum states is irreversible FAPP, but we will not analyse them here further. 

We conclude that the lack of the complete knowledge of the Hamiltonian circumvents the time reversal objection, also known as Loschmidt's paradox, which states that it should not be possible to deduce time irreversibility from an underlying time reversal theory. A similar argument could be applied to address the recurrence objection, which is based on the Poincar{\' e} recurrence theorem, that all finite systems are recurrent, i.e. return arbitrarily close to their initial state after a possibly very long time. Results show that recurrence times in the dynamics of quantum states could be extremely large~\cite{peres4}. A comprehensive study of various models of quantum measurement can be found in Ref.~\cite{balian}

\subsection*{Deutsch's thought experiment}

In Ref.~\cite{deutsch}, Deutsch proposed an experiment which he claims can distinguish experimentally between the Copenhagen and the Everett interpretations of quantum mechanics. While I do not in any way see the necessity of assuming that the two interpretations might have distinct predictions in the experiment, I acknowledge that the experiment most strikingly demonstrates the necessity of a radical revision of our attitude to physical reality in quantum physics.

\begin{figure}
\begin{center}
\includegraphics[width=12.4cm]{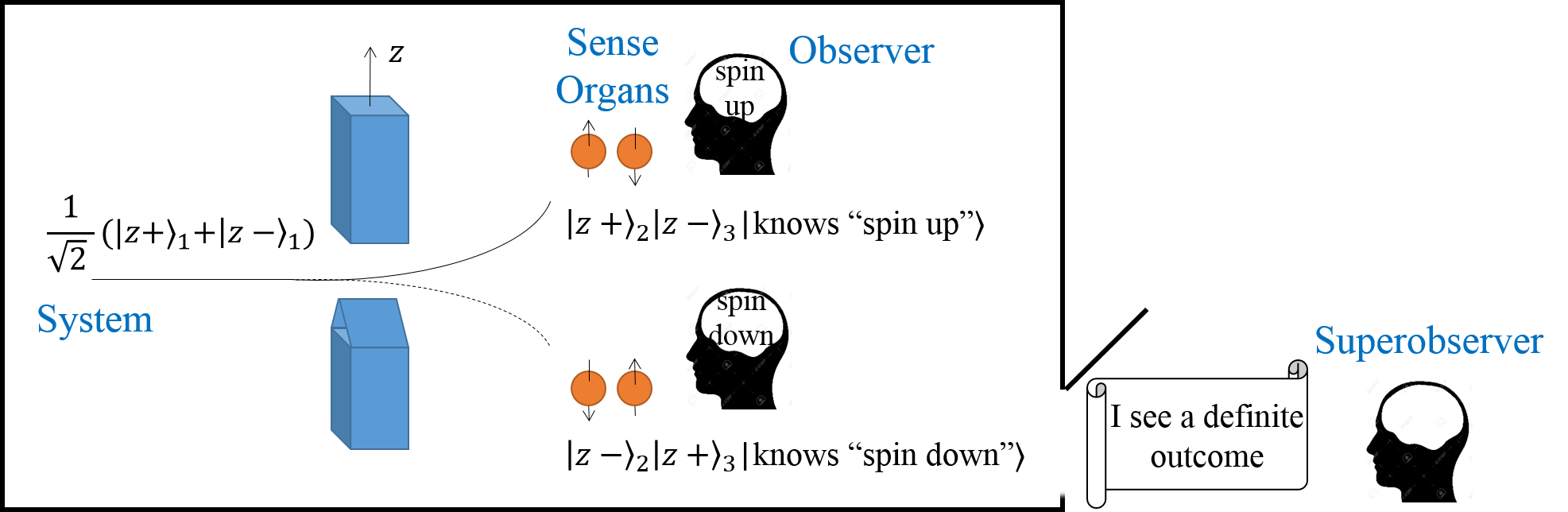}
\end{center}
\caption{Deutsch's version of the Wigner's friend gedanken experiment. An observer performs the Stern-Gerlach experiment on a spin-1/2 atom (system 1) in a closed laboratory. The outcome ``spin up'' or ``spin-down'' is recorded in sense organs, which are also spin-1/2 atoms (systems 2 and 3,) and finally in the observer's brain (system 4). The outside observer, the superobserver, describes the experiment as a coherent evolution of a large entangled state. The observer communicates a message to the superobserver outside, which contains information about whether she sees a definite outcome or not, without revealing which outcome she sees. What will the observer then experience? Would the superobserver in principle be able to perform an interference experiment on the systems and the observer and confirm the appropriateness of his state assignment?} \label{figure2}
\end{figure}

The thought experiment involves measurements on the observer by another, superobserver, and is a variant of the Wigner's-friend thought experiment~\cite{wigner1}. Four systems are involved in the experiment as illustrated in Fig.~\ref{figure2}. System 1 is a spin-1/2 atom which passes through a Stern-Gerlach apparatus in such a way that the two trajectories, corresponding to outcomes ``spin up'' and ``spin down'', pass over systems 2 and 3. These two systems, also spin-1/2 atoms, represent part of the observer's ``sense organ''. Their receptive states at the outset are ``spin down''. They are coupled with atom 1 in such away that when  atom 1 follows the ``spin up'' trajectory, it passes over atom 2. The spin of atom 2 now flips to ``spin up''. Meanwhile, the spin of atom 3 remains the same. If in a similar way atom 1 follows the ``spin-down'' trajectory, the spin of atom 3 will flip while the spin of atom 2 will remain unchanged. System 4 is the observer, and it couples only to sense organs 2 and 3. Potentially, there are further systems that constitute an environment of the four systems. They all are isolated from the rest in a sealed laboratory. The experiment is then performed a sufficient number of times to collect statistics.

Initially, the state of the four systems is factorized with atom 1 in state $|x+\rangle_1 = \frac{1}{\sqrt{2}} (|z+\rangle_1 + |z-\rangle_1)$ and the observer in some definite state, $|0\rangle_4$, whose exact properties do not need to be specified, except that she is capable of completing a measurement:
\begin{equation}
|\psi(0)\rangle = \frac{1}{\sqrt{2}} (|z+\rangle_1+|z-\rangle_1)|z-\rangle_2|z-\rangle_3|0\rangle_4.
\end{equation}
One can also consider mixed states, but this assumption would complicate the situation unnecessarily. The Stern-Gerlach magnet is assumed to be oriented along the $z$-axis. After atom 1 has passed through the Stern-Gerlach apparatus and has interacted with the sense organs at time $t$, the state is 
\begin{equation}
|\psi(t)\rangle = \frac{1}{\sqrt{2}} (|z+\rangle_1|z+\rangle_2|z-\rangle_3 +|z-\rangle_1|z-\rangle_2|z+\rangle_3 )|0\rangle_4.
\end{equation}
Finally, after the interaction between the observer and the sense organs at time $t'$, the state becomes
\begin{equation}
|\psi(t')\rangle = \frac{1}{\sqrt{2}} (|z+\rangle_1|z+\rangle_2|z-\rangle_3|\mbox{knows ``up''}\rangle_4 +|z-\rangle_1|z-\rangle_2|z+\rangle_3 |\mbox{knows ``down''}\rangle_4),\label{super}
\end{equation}
where $|\mbox{knows ``up''}\rangle_4$ and $|\mbox{knows ``down''}\rangle_4$ denote the observer's states after recording the result. If there are further systems in the laboratory, their states can eventually also get correlated with the two amplitudes in Eq.~(\ref{super}) in a huge entangled state. Unless stated otherwise, I will assume that there are no further systems in the laboratory. 

Strictly speaking, the quantum state~(\ref{super}) can have an operational meaning only for the superobserver, who is stationed outside the sealed laboratory. To him on the outside, on the basis of all the information that is in principle available to him and conditioned on having sufficient experimental capabilities, the physical description of the state in the laboratory will be the superposition~(\ref{super}). He can test the validity of the state assignment by performing an interference experiment with the output states:
\begin{eqnarray}
|\psi+\rangle &=&\frac{1}{\sqrt{2}} (|z+\rangle_1|z+\rangle_2|z-\rangle_3|\mbox{knows ``up''}\rangle_4 +|z-\rangle_1|z-\rangle_2|z+\rangle_3 |\mbox{knows ``down''}\rangle_4) \nonumber \\
|\psi- \rangle &=& \frac{1}{\sqrt{2}} (|z+\rangle_1|z+\rangle_2|z-\rangle_3|\mbox{knows ``up''}\rangle_4 - |z-\rangle_1|z-\rangle_2|z+\rangle_3 |\mbox{knows ``down''}\rangle_4). \nonumber
\end{eqnarray}
This requires a special experimental arrangement and instruments of high measurement precision, which allow measuring the systems' and the observer's brain states in coherent superpositions.

What will the observer in the laboratory perceive in state~(\ref{super}) after completing of her measurement? Will she definitively know if she has observed one single outcome or not? It is tempting to answer such questions within the standard quantum framework: Within the laboratory, the actual observation projects the quantum state into one of the two possibilities. The observer will therefore either observe outcome ``spin up'' or outcome ``spin-down''. We know that for the projection to occur, it is not necessary for the observer to actually read out the information from the measurement device; it is sufficient that the information is available in principle~\cite{mandel}. Then, if the superobserver were to project his state onto the basis of ``all information that is in principle available'' to him, would that not include information that is available to the observer? Should the mere availability of the information about the outcome somewhere -- specifically, in the observer's brain -- not collapse the quantum state that the superobserver assigns? Or does the observer observe some kind of ``blurred reality'', while the superobserver keeps describing the situation in terms of the superposition state? Deutsch's ingeniously contrived experiment could answer these questions at least in principle, albeit its execution is impractical. 

The idea is that the superobserver could learn whether the observer has observed a definite outcome, without himself learning which outcome she has observed. It is enough for the observer to communicate ``I observe a definite outcome'' or ``I observe no definite outcome'' to the superobserver. (For this purpose, the laboratory may be opened to pass only this message, keeping all other degrees of freedom still fully isolated. While being practically demanding, this is possible in principle.) The message could, for example, be written on a piece of paper and passed on to the superobserver. The key element of the experiment is that the message contains no information about which outcome has occurred and thus should not lead to a collapse of the quantum state assigned by the superobserver. Imagine that the observer encodes her message in state $|\mbox{message}\rangle_5$ of system 5. This state is factorized out from the total state, $|\psi(t')\rangle = \frac{1}{\sqrt{2}} (|z+\rangle_1|z+\rangle_2|z-\rangle_3|\mbox{knows ``up''}\rangle_4 +|z-\rangle_1|z-\rangle_2|z+\rangle_3 |\mbox{knows ``down''}\rangle_4)$ $|\mbox{message}\rangle_5$, and thus the communication of the message does not destroy the superposition.

What will be written in the message? Will the superobserver see the interference? Three different results of the experiment are possible\footnote{In a quantum mechanical experiment, the ``observer'' could be simulated by a qutrit with the following encoding~\cite{bennett}: $|0\rangle$ for ``knowing spin up'', $|1\rangle$ for ``knowing spin down'' and $|2\rangle$ for ``I see no definite outcome''. The message is then encoded either in $|2\rangle$ or in a state with the two-dimensional support spanned by vectors $|0\rangle$ and $|1\rangle$ (for example $\frac{1}{2}(|0\rangle\langle 0|+|1\rangle \langle 1|$). The superobserver applies the measurement with the projectors $\hat{P}_1=|0\rangle \langle 0| + |1\rangle \langle 1|$ and $\hat{P}_2=|2\rangle \langle 2|$.}:

\begin{enumerate}

\item The quantum state collapses due to a breakdown of the quantum-mechanical laws when applied to states of brain or to systems of sufficiently large size, mass, complexity, and the like. The collapse models Ghirardi-Rimini-Weber~\cite{GRW} or Diosi-Penrose~\cite{diosi,penrose} fall into this category. One could also argue in favor of the collapse within the view according to which a quantum state is a representation of the observer's knowledge. Every measurement yields new information, and the representation of this knowledge update is the state projection. Since the new information about the outcome is available somewhere -- specifically in the observer's brain -- the state has to collapse for {\it all} observers, including the superobserver\footnote{It seems to me that Deutsch had this particular view in mind when he claimed that the Copenhagen interpretation predicts the occurrence of the collapse. I see this view at most as a variant of the interpretation and (to my knowledge) not widely spread.}. Independently of the specific rationale behind the state collapse, the observer sends the message that she observers a definite outcome. The superobserver concludes that although he could exclude all known effects caused by conventional decoherence, the state is not in the superposition. This he can confirm in the interference experiment by observing that both outputs in the interference experiments occur with equal probability.  

\item The superobserver's state assignment is the superposition state, and the observer perceives a ``blurred reality'' that she associates with not seeing a definite outcome. She sends a message: ``I observe no definite outcome''. The superobserver confirms the superposition state in the interference experiment by observing a single output state in the interference experiment. I personally have trouble to make sense of this option. If quantum theory describes an  observer's probability assignments in well-defined experimental procedures, where, to quote Bohr~\cite{bohr2} ``... by the word 'experiment' we refer to a situation where we can tell others what we have done and what we have learned ...'', then experience of ``blurred reality'' seems to be outside of the standard quantum framework. Moreover, such a situation would install a fundamental asymmetry between the observers, those who see and those who do not see ``blurred reality''.

\item The quantum laws are unmodified. The superobserver's state assignment is the superposition state. And yet, the observer observes a definite outcome. The assigned superposition state can be confirmed in the interference experiment. 

\end{enumerate}

In my eyes, outcomes 1 and 2 would indicate fundamentally new physics. I will not consider these cases further and regard quantum theory to be a {\it universal} physical theory. This leaves us with situation 3 as the only possible outcome of Deutsch's thought experiment. The outcome is compatible with the Everett interpretation: each copy of the observer observes a definite but different outcome in different branches of the (multi)universe. The outcome is compatible with the Copenhagen interpretation too, but it is rarely discussed what the implications of this claim are for our understanding of physical reality within the interpretation. The rest of the current manuscript is devoted to this problem.   

Note that in situation 3 of the thought experiment, the two observers have complementary pieces of information. {\it Taken together}, they would violate the complementarity principle of quantum physics. The observer has complete knowledge about the value of observable $A_1$ with eigenstates $|z+\rangle_1|z+\rangle_2|z-\rangle_3|\mbox{knows ``up''}\rangle_4$ and $|z-\rangle_1|z-\rangle_2|z+\rangle_3 $ $ |\mbox{knows ``down''}\rangle_4$, whereas the superobserver has complete knowledge about the value of observable $A_2$ with eigenstates $|\psi+\rangle$ and $|\psi-\rangle$. The two observables are non-commuting. One might be tempted to interpret outcome 3 of Deutsch's experiment as implying that the two pieces of information coexist. After all, the superobserver has evidence -- in form of the message -- that the observer had perfect knowledge about $A_1$. And yet, on the very same state~(\ref{super}), he can learn the value of $A_2$. Even the observer herself, retrospectively, after completion of the interference experiment, can be convinced that there is a discrepancy between her message and the fact that she always ends up in one output state in the interference experiment (thereby forgetting which outcomes she had observed). This is because, if she previously were in a state observing a definite outcome, then by applying standard quantum mechanical predictions on the systems and herself (which in itself is a problematic step because it ignores the necessity of the object-subject cut), she should have equal probability to end up in either of the two output states. 

The trouble with the assumption that values for $A_1$ and $A_2$ {\it coexist} is that it introduces ``hidden variables'', for which a Bell's theorem can be formulated with its known consequences. To this end, consider a pair of superobservers, Alice and Bob, who reside in their local laboratories and conduct an experiment involving observers Anton and Bertlmann, respectively, who in turn each perform a Stern-Gerlach experiment of the type explained above. More specifically, Alice has four systems in her laboratory: atom A1, sense organs A2 and A3, and observer Anton A4. Similarly, Bob has four systems: atom B1, sense organs B2 and B3, and observer Bertlmann B4. Suppose that the two superobservers share an entangled state:

\begin{equation}
|\psi\rangle_{AB} = \frac{1}{\sqrt{2}} (|A_{up}\rangle|B_{down}\rangle - |A_{down}\rangle |B_{up}\rangle), \label{bell}
\end{equation}
where 
\begin{eqnarray}
|A_{up}\rangle &=& |z+\rangle_{A1} |z+\rangle_{A2}|z-\rangle_{A3}|\mbox{Anton knows ``up''}\rangle_{A4}, \nonumber  \\
|A_{down}\rangle &=&|z-\rangle_{A1}|z-\rangle_{A2}|z+\rangle_{A3} |\mbox{Anton knows ``down''}\rangle_{A4}, \nonumber \\
|B_{up}\rangle &=& |z+\rangle_{B1}|z+\rangle_{B2}|z-\rangle_{B3}|\mbox{Bertlmann knows ``up''}\rangle_{B4}, \nonumber \\
|B_{down}\rangle&=& |z-\rangle_{B1}|z-\rangle_{B2}|z+\rangle_{B3} |\mbox{Bertlmann knows ``down''}\rangle_{B4}. \nonumber
\end{eqnarray}
Using these states, one can define observables that are analogues to spin projections along the $z$ and $x$ axes of a spin-1/2 particle, respectively: $A_z=|A_{up}\rangle \langle A_{up}| - |A_{down} \rangle \langle A_{down}|$, $A_x= |A_{up}\rangle \langle A_{down}| + |A_{down} \rangle \langle A_{up}|$ for Alice, and similarly for Bob. Note that eigenstates of $A_z$ correspond to the observer's states ``knowing the spin-{\it z} to be up'' and ``knowing the spin-{\it z} to be down'', and those of $A_x$ to the possible outcomes of the superobserver's interference experiment.

\begin{figure}
\begin{center}
\includegraphics[width=14.2cm]{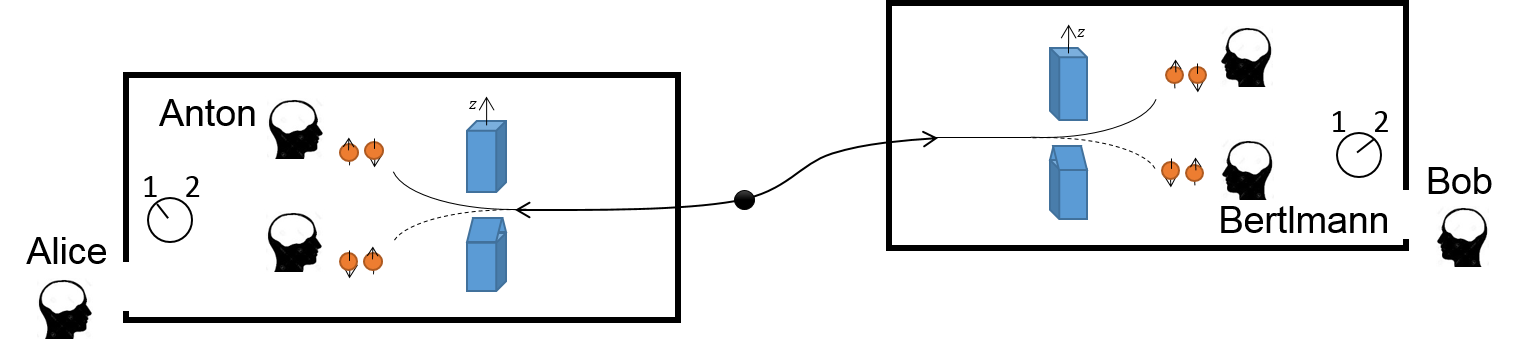}
\end{center}
\caption{Bell's experiment to exclude the coexistence of ``facts'' (i.e. measurement outcomes or records) for both the observer and the superobserver. Alice and Bob (both of them superobservers) reside in their space-like separated laboratories in which two further observers, Anton and Bertlmann respectively, perform a Stern-Gerlach type of measurement. By choice of local measurement setting (1 or 2), each of the superobservers can either interrogate which outcome the respective observer in his laboratory has observed or perform the interference experiment jointly on the observer and the spin. With a suitable entangled state~\ref{bell}, the superobservers can violate Bell's inequality.} \label{figure3}
\end{figure}

Let us assume that in the Bell experiment, Alice chooses between two measurement settings $A_1=A_z$ and $A_2= A_x$ and Bob between $B_1= \frac{1}{\sqrt{2}} (B_z + B_x)$ and $B_2 =\frac{1}{\sqrt{2}} (B_z - B_x)$. In a local (deterministic) hidden variable theory, one assumes that there jointly exist predetermined values for $A_1$, $A_2$, $B_1$ and $B_2$ which are +1 or -1.  It is a well-known fact that state~(\ref{bell}) with the chosen settings leads to a violation of the Bell-Clauser-Horne-Shimony-Holt inequality $|\langle A_1 B_1 \rangle + \langle A_1 B_2 \rangle + \langle A_2 B_1 \rangle - \langle A_2 B_2 \rangle| \leq 2$, where 
$\langle A_i B_j \rangle$, $i,j=1,2$, is the correlation function. The maximal quantum value for the Bell expression is $2\sqrt{2}$. Just like in every other Bell test, we conclude that the definite values for the observables cannot coexist if one keeps the assumption of locality\footnote{Here ``locality'' means that, for example, value $A_1$ depends only on the local setting of Alice and not on the distant one of Bob. In a non-local hidden variable theory, we would need to distinguish between $A_{11}$ and $A_{12}$, depending on whether Bob's setting is 1 or 2, respectively. It is not necessary to assume local deterministic values to derive Bell' inequalities. Bell's local causality is sufficient~\cite{bell}. This however does not change the conclusions~\cite{marek}.}. We conclude that the two pieces of information, one of the observer and another of the superobserver, cannot be taken to coexist. 

What consequences does outcome 3 of Deutsch's thought experiment have for our understanding of physical reality? Let us assume that the observers' and superobservers' laboratories contain a large number of degrees of freedom which allow the information about respective measurement records  to be FAPP redundantly imprinted in their respective ``environments''. I will call these records ``facts''. This could be a click in a photodetector, a certain position of a pointer device, a printout of a computer or a written page in the lab-book, or a definite human brain state of a colleague who read the lab-book. If we assume that all these records in the observer's laboratory get correlated with the spin atoms and her brain state, {\it and} the superobserver can still perform the interference experiment, the result of which is also recorded in his laboratory, one has to accept that the two pieces of information can redundantly be imprinted in {\it two} environments: the sealed laboratory and the outside, respectively.  As long as there is no communication on the relevant information (the actual measurement outcome) between the two laboratories, they will remain separate. 

If we respect that there should be no preferred observers, then there is no reason to assume that the ``facts'' of one of them are more fundamental than those of the other\footnote{One might be tempted to assume that the ``facts'' of the superobserver are the ``real'' ones, as he definitely has more reliable measurement instruments than the observer. This view cannot withstand the objection that the superobserver himself might be an object observed by yet another observer, the supersuperobserver, who describes the interference experiment of the superobserver quantum mechanically. The regression of increasingly more powerful observers might eventually find its end in a universe with a finite amount of resources.}. But then, the observers' records cannot be comprised as ``facts of the world'', independent of the ``environment'' in which they have occurred. Any attempt to introduce ``facts of the world {\it per se}'' would run into problems of the hidden variable program. 

The implications of the present Bell experiment are stronger than those of the standard Bell test. In the latter, we can exclude the view according to which the outcomes for measurements are (locally) predetermined, no matter if any measurement -- and no matter which measurement -- is actually performed. Still, between the partners there is no ambiguity with respect to whether measurements take place and about the coexistence of their records. The records can be accomplished as ``facts of the world'', which they share and even need to communicate in order to evaluate the experimental bound of the Bell expression. This is no longer the case in the present Bell experiment. What the Bell experiment excludes is the coexistence of the ``facts'' themselves. Everettians solve this by assuming that mutually complementary facts never coexist in between two branchings of the (multi)universe. Copenhagenists (can) take the position that {\it there are no facts of the world per se, but only relative to observers}. This is similar to Quantum Bayesianism, which treats the state of a quantum system as being observer-dependent, and to Rovelli's relational quantum mechanics~\cite{rovelli}, according to which ``quantum mechanics is a theory about the physical description of physical systems relative to other systems ...'' There are, however, important differences.  

In Rovelli's relational interpretation, the ``observer'' does not ``make any reference to a conscious, animate, or computing, or in any other manner special, system''~\cite{rovelli} -- each system provides its own frame of reference relative to which states of other systems can be assigned. Taking this position and outcome 3 of the Deutsch's experiment and applying them to, for example, the interference phenomenon in the double-slit experiment with single electrons, one would conclude that, although the observer has no path information, the electron itself ``knows'' which path it takes. Relative to the electron, a definite path is taken, although we as observers observe an interference pattern. Obviously, we are here encountering the limits of meaningful language when we associate the terms ``knowledge'' or ``taken'' to single electrons. In this respect, quantum theory (in my eyes) remains a fundamental theory of observations in which a (hypothetical) observer, measurement and probabilities play a central role.

The two dominant approaches to the probability interpretation are the frequentist approach and the Bayesian approach. Qbism 
views the quantum state to be a user's manual --  a mathematical tool that an observer uses to make decisions and take actions on the surrounding world upon observations. Central to this position is a Bayesian or personalist probabilistic approach to the probabilities that appear in quantum theory. To me, however, the problem of probability interpretation is prior to quantum theory, the solution of which alone will not be able to answer the question: What are the invariant features that characterize quantum theory in ways that are not relative to observers? By taking the subject matter of quantum theory to be restricted to an individual agent's decisions and experiences, Qbism runs into the danger of denying any objective elements in the notion of the quantum state. I agree with the Qbist's notion of subjective quantum states as representatives of an agent's beliefs, but only to the extent where a fundamental limit on maximal possible degree of belief of any agent is respected. This limit is represented FAPP by a pure quantum state. The fact that the predictions of agents cannot be ``improved'' over and above this limit in my eyes indicates that probabilities are not just personal and subjective, but also formed by the aforementioned invariant features of the theory. The role of reconstructions of quantum theory is to identify these invariants\footnote{In recent years, there have been several attempts to account for the origin of the basic principles from which the structure of quantum theory can be derived without invoking mathematical terms such as ``rays in Hilbert space'' or ``self-adjoint operators.''~\cite{hardy,dakicbrukner,masanesmueller,chiribella}}.

The difference to the Everett interpretation is more evident. In the view adopted here, no meaning is given to ``the universal wave function'', nor is there an attempt to arrive at the probabilities from within such a concept alone. Here, the probabilities are always given by the Born rule, which is part of the formalism. This applies also to superobservers of any order: probabilities acquire meaning only when the measurement arrangement is specified, in which these probabilities are observed. 

Finally, I comment on the view~\cite{grangier} that the cut cannot be moved to include measurement instruments, observers etc. as objects under observation, since an object can never grow up to the point that it includes measurement contexts that, in turn, are unavoidably given in terms of classical concepts in accordance to Bohr's doctrine~\cite{bohr1}: ``However far the (quantum) phenomena transcend the scope of classical physical explanation, the account of all evidence must be expressed in classical terms.'' According to this view~\cite{grangier}, the necessity of unambiguous usage of classical concepts fixes the object-subject cut whose position is therefore fundamental and equal for {\it all} observers. Consequently, one can retain the objectivity of the ``facts of the world''. I do not think that this view stands up to closer scrutiny. The description of any quantum mechanical experiment is expressed ``in common language supplemented with the terminology of classical physics''~\cite{bohr4}. Although this observation has played an important role in clarifying misconceptions in debates over the interpretation of quantum theory, it is in retrospective rather self-evident. For example, the description of a double-slit experiment with atoms, includes the depiction of the source of atoms directed towards the diaphragm normal to the beam, where the diaphragm contains two slits and a photographic plate with a characteristic interference pattern on the plate where the atoms are deposited. By extending the experiment to larger and larger systems, eventually as large as measurement instruments, nothing should change in the epistemic basis of the theory: we will still give an unambiguous account of the phenomenon in terms of classical language including a suitable ``source'', ``beam'' and ``observation screen''. This should not be confused with the impossibility of giving a classical explanation of the phenomenon, e.g. in terms of  well-defined classical trajectories, which is present both for atoms and for macroscopic objects. To conclude, the cut can be shifted with no change in the epistemic foundation of the theory. Negating this would either mean negating Wigner-type experiments as legitimate quantum mechanical experiments or predicting outcome 1 in Deutsch's experiment. Both choices indicate an acceptance that quantum theory is not universal. 

The above-mentioned Bell's theorem for ``facts'' implies a striking departure from naive realism. This brings us to the question of the role of our physical theories. If physical theories do not describe ``physical reality {\it per se}'', what do they describe then? A possible answer is given by Bohr as communicated by Petersen~\cite{bohr}:``It is wrong to think that the task of physics is to find out how nature is. Physics concerns what we can say about nature''.

\subsection*{Acknowledgements}
This work has been supported by the Austrian Science Fund (FWF) through CoQuS, SFB FoQuS, and Individual Project 2462. I would like to acknowledge discussions with Mateus Araujo, Borivoje Daki{\'c}, Philippe Grangier, Richard Healey, Johannes Kofler, Luis Masanes and Anton Zeilinger.

%
%
%


\begin {thebibliography}{100}

\bibitem{bub} J. Bub and I. Pitowsky, {\it Two dogmas about quantum mechanics}, in Many Worlds? Everett, Quantum Theory, and Reality, S. Saunders, J. Barrett, A. Kent, and D. Wallace (eds.), pp. 431-456 (Oxford University Press, 2010).

\bibitem{pitowsky} I. Pitowsky, {\it Quantum mechanics as a theory of probability}, in Festschrift in honor of Jeffrey Bub, W. Demopoulos and I. Pitowsky (eds.) (Springer, Western Ontario Series in Philosophy of Science, New York, 2007). 

\bibitem{maudlin} T. Maudlin, {\it Three measurement problems}, Topoi {\bf 14} (1), 7-15 (1995).

\bibitem{cbaz} C. Brukner and A. Zeilinger, {\it Information and fundamental elements of the structure of quantum theory}, in ``Time, Quantum, Information'', edited by L. Castell and O. Ischebeck (Springer, 2003).

\bibitem{qbism} C. A. Fuch and R. Schack, {\it Quantum-Bayesian coherence}, Rev. Mod. Phys. {\bf 85}, 1693 (2007).

\bibitem{rovelli} C. Rovelli, {\it Relational quantum mechanics}, Int. J. Th. Phys. {\bf  35}, 1637-1678 (1996).

\bibitem{colbeck} R. Colbeck and R. Renner, {\it No extension of quantum theory can have improved predictive power}, Nature Communications {\bf 2}, 411 (2011).

\bibitem{GRW} G.C. Ghirardi, A. Rimini, and T. Weber, {\it Unified dynamics for microscopic and macroscopic systems}, Phys. Rev. D {\bf 34}: 470 (1986).

\bibitem{diosi} L. Diosi, {\it Models for universal reduction of macroscopic quantum fuctuations}, Phys. Rev. A {\bf 40}, 1165-1174
(1989).

\bibitem{penrose} R. Penrose, {\it On gravity's role in quantum state reduction}, Gen. Relat. Gravit. {\bf 28}, 581-600 (1996).

\bibitem{bell} J. S. Bell, {\it Speakable and unspeakable in quantum mechanics}, Collected papers on quantum philosophy (Cambridge Univ. Press, 2004).

\bibitem{marek} M. Zukowski and  {\v C}. Brukner, {\it Quantum non-locality - it ain't necessarily so ...}, Special issue on 50 years of Bell's theorem, J. Phys. A: Math.Theor. {\bf 47}, 424009 (2014).

\bibitem{fuchs} C. A. Fuchs and R. Schack, {\it QBism and the Greeks: why a quantum state does not represent an element of physical reality}, arXiv:1412.4211 (2014). 

\bibitem{osnaghi} S. Osnaghi, F. Freitas, O. Freire Jr., {\it The origin of the Everettian heresy}, Stud. Hist. Philos. Mod. Phys. {\bf 40} (2), 97-123 (2009).

\bibitem{max} K. Camilleri and M. Schlosshauer, {\it Niels Bohr as philosopher of experiment: Does decoherence theory challenge Bohr's doctrine of classical concepts?}, Stud. Hist. Phil. Mod. Phys. {\bf 49}, 73-83 (2015).

\bibitem{heelan} P. Heelan, {\it Heisenberg and radical theoretical change}, Z. Allgemeine Wissenschaftstheorie {\bf 6}, 113-138 (1975).

\bibitem{AHQP} AHQP, 1986. Archives for the History of Quantum Physics -- Bohr's Scientific Correspondence, American Philosophical Society, Philadelphia, 301 microfilm reels.

\bibitem{spekkens} N. Harrigan and R. W. Spekkens, {\it Einstein, incompleteness, and the epistemic view of quantum states}, Found. Phys. {\bf 40}, 125 (2010).

\bibitem{malin} S. Malin, {\it What are quantum states?}, Quantum Information Processing, {\bf 5}, 233-237 (2006).

\bibitem{wigner} E. P. Wigner, {\it The problem of measurement}, Am. J. Phys. {\bf 31}, 6 (1963).

\bibitem{peres} A. Peres, {\it When is a quantum measurement?}, Annals of the New York Academy of Sciences {\bf 480}, New Techniques and Ideas in Quantum Measurement Theory, 438 (1986).

\bibitem{wigner} E. Wigner, Symmetries and Reflections, p. 164 (Indiana University Press, 1967).

\bibitem{heisenberg} W. Pauli, Wissenschaftlicher Briefwechsel mit Bohr, Einstein, Heisenberg, Vol. 2, edited by K. von Meyenn, A. Hermann, and V. F. Weisskopf, (Springer, Berlin, 1985), pp. 1930-1939.
For the English translation of Heisenberg's manuscript with an introduction and bibliography see E. Crull and G. Bacciagaluppi, http://philsci-archive.pitt.edu/8590/.

\bibitem{mueller} M. P. Mueller and L. Masanes, {\it 
Three-dimensionality of space and the quantum bit: an information-theoretic approach}, New J. Phys. {\bf 15}, 053040 (2013).

\bibitem{dakic} B. Dakic and {\v C}. Brukner, {\it The classical limit of a physical theory and the dimensionality of space}, to appear in ``Quantum Theory: Informational Foundations and Foils'', G. Chiribella, and R. Spekkens; (eds.)  Preprint at arXiv:1307.3984.

\bibitem{chiribella} G. Chiribella, G. D'Ariano, and P. Perinotti, {\it Informational derivation of quantum theory}, Phys. Rev. A {\bf 84}, 012311 (2011).

\bibitem{kofler} J. Kofler and {\v C}. Brukner, {\it Classical world arising out of quantum physics under the restriction of coarse-grained measurements}, Phys. Rev. Lett. {\bf 99}, 180403 (2007).

\bibitem{zurek} W. H. Zurek, {\it Decoherence, einselection, and the quantum origins of the classical}, Rev. Mod. Phys. {\bf 75}, 715 (2003).


\bibitem{kofler2} J. Kofler and {\v C}. Brukner, {\it Conditions for quantum violation of macroscopic realism}, Phys. Rev. Lett. {\bf 101}, 090403 (2008).

\bibitem{peres2} A. Peres, Quantum Theory: Concepts and Methods (Kluwer Academic Publishers, New York, 1995).

\bibitem{peres3} A. Peres, {\it Stability of quantum motion in chaotic and regular systems}, Phys. Rev. A {\bf 30}, 1610 (1984).

\bibitem{chaos} Hans-J{\"u}rgen St{\"o}ckmann, Quantum Chaos: An Introduction (Cambridge University Press, Cambridge, 1999).

\bibitem{jacquod} Ph. Jacquod, I. Adagideli, C.W. J. Beenakker, {\it Decay of the Loschmidt echo for quantum states with sub-Planck-scale structures}, Phys. Rev. Lett. {\bf 89}, 154103 (2012). 

\bibitem{peres4} A. Peres, {\it Recurrence phenomena in quantum dynamics}, Phys. Rev. Lett. {\bf 49}, 1118 (1982).

\bibitem{balian} A. E. Allahverdyan, R. Balian, T. M. Nieuwenhuizen,
{\it Understanding quantum measurement from the solution of dynamical models}, Physics Reports {\bf 525}, 1 (April 2013)

\bibitem{deutsch} D. Deutsch, {\it Quantum theory as a universal physical theory}, Int. J. Th. Phys. {\bf 24}, I  (1985).

\bibitem{wigner1} E. P. Wigner, {\it Remarks on the mind-body question}, in: I. J. Good, ``The Scientist Speculates'' (London, Heinemann, 1961)

\bibitem{mandel} X. Y. Zou, T. P. Grayson, and L. Mandel, {\it Observation of quantum interference effects in the frequency domain}, Phys. Rev. Lett. {\bf 69}, 3041 (1992).

\bibitem{bohr1} N. Bohr, The philosophical writings of Niels Bohr 3 (Woodbridge, Conn.: Ox Bow Press, 1987).

\bibitem{bennett} C. Bennett, private communication.

\bibitem{bohr2} N. Bohr, {\it Discussion with Einstein on epistemological problems in atomic physics}, in Albert Einstein: Philosopher-Scientist, Ed. Paul Arthur Schilpp, (Evanston, Illinois: The Library of Living Philosophers, 1949).

\bibitem{hardy} L. Hardy, Quantum theory from five reasonable axioms, Preprint at arXiv:quant-ph/0101012 (2001).

\bibitem{dakicbrukner} B. Dakic and C. Brukner, {\it Quantum theory and beyond: Is entanglement special?}, in ``Deep Beauty: Understanding the Quantum World through Mathematical Innovation'', Ed. H. Halvorson, (Cambridge University Press, 2011), 365-392.

\bibitem{masanesmueller} L. Masanes and M. M{\" u}ller, {\it A derivation of quantum theory from physical requirements}, New J. Phys. {\bf 13}, 063001 (2011).

\bibitem{grangier} A. Auff{\ 'e}ves, and P. Grangier, {\it Contexts, systems and modalities: a new ontology for quantum mechanics}, Preprint at arXiv:1409.2120 (2014).

\bibitem{bohr4} N. Bohr, {\it On the notions of causality and complementarity}, Dialectica {\bf 2}, 312-319 (1948).

\bibitem{bohr} As quoted in ``The philosophy of Niels Bohr'' by Aage Petersen, in the Bulletin of the Atomic Scientists {\bf 19}, No. 7 (September 1963); ``The Genius of Science: A Portrait Gallery`'' (2000) by Abraham Pais, p. 24, and ``Niels Bohr: Reflections on Subject and Object'' (2001) by Paul. McEvoy, p. 291.

%
%
%
%
%
%
%

\end{thebibliography}
\end{document}